\documentclass{JHEP3}
\usepackage{amsmath,amssymb,graphicx}
\newcommand{\ket}[1]{\ensuremath{| #1 \rangle}}
\newcommand{\bra}[1]{\ensuremath{\langle #1 |}}
\newcommand{\trace}{\text{Tr}}
\title{Entanglement Interpretation of Black Hole Entropy \\ in String Theory.}
\author{Ram Brustein${}^a$, Martin B. Einhorn${}^{b,c}$, Amos Yarom${}^a$ \\
${}^a$Department of Physics, Ben-Gurion University, Beer-Sheva 84105, Israel \\
${}^b$Kavli Institute for Theoretical Physics, University of California, Santa
Barbara, CA 93106
\\ ${}^c$Michigan Center for Theoretical Physics, Randall Laboratory, The
University of Michigan,
Ann Arbor, MI 48109 \\
    {\rm E-mail:}
    {\tt ramyb@bgu.ac.il, meinhorn@kitp.ucsb.edu, yarom@bgu.ac.il}}

\abstract{ We show that the entropy resulting from the counting of microstates of
non extremal black holes using field theory duals of string theories can be
interpreted as arising from entanglement. The conditions for making such an
interpretation consistent are discussed. First, we interpret the entropy (and
thermodynamics) of spacetimes with non degenerate, bifurcating Killing horizons as
arising from entanglement.  We use a path integral method to define the
Hartle-Hawking vacuum state in such spacetimes and discuss explicitly its entangled
nature and its relation to the geometry.   If string theory on such spacetimes has
a field theory dual, then, in the low-energy, weak coupling limit, the field theory
state that is dual to the Hartle-Hawking state is a thermofield double state. This
allows the comparison of the entanglement entropy with the entropy of the field theory
dual, and thus, with the Bekenstein-Hawking entropy of the black hole. As an example,
we discuss in detail the case of the five dimensional anti-de~Sitter, black hole
spacetime.}

\begin{document}


\maketitle
\section{Introduction}
The discovery that black holes have entropy \cite{Bekenstein} and
that they radiate \cite{Hawking}, has led to many speculations about
their quantum gravity origin. The unusual non-extensive nature of
the entropy of black holes and their effect on matter thrown into
them, leads to conjectures that entropy bounds should exist
\cite{BekEBounds}. The area scaling character of the entropy seems
to imply a ``holographic principle''\cite{HoltHooft,HolSusskind},
where a system has an areas' worth of physical degrees of freedom
\cite{Bousso}. The celebrated AdS/CFT correspondence
\cite{AdSCFTreview} is one such example. Yet, about 30 years after
its discovery, solid information regarding the physical origin of
black hole entropy is still lacking.

The entropy of a thermodynamic system is a measure of the number of
the available microstates of the system that can not be
discriminated by an observer who measures macroscopic quantities.
For example, when considering black body radiation, one needs to
count the number of the appropriate photon configurations. The
entropy of this ``photon gas'' is the logarithm of the number of
states in phase space that are available to the photons. From this
perspective black hole entropy relies on some fundamental definition
of microstates in a theory of quantum gravity: it is a measure of
the number of microstates of the black hole that an observer would
not be aware of when measuring macroscopic parameters.

String theory has been able to successfully count these microstates explicitly
\cite{StromVafa, MalStrWit, Bring, Peetetal}. This has been done by appealing to a
duality between gravitational systems and (conformal) field theories. Black holes
are described in a low energy limit of a gravitating theory. They also have an
equivalent description as a field theory, whereby their microstates may be counted
explicitly. This method does not give physical insight regarding the nature of the
microstates of the black hole nor does it offer a reason for its area scaling
property. Recent work regarding counting of microstates from a different
perspective can be found in \cite{SenBHstring,Mathur,Mathur:2004sv}.

The idea that we pursue is that the microstates are those due to
entanglement of the vacuum of the black hole.  Its main features are the
observer-dependence of the entropy, the role of quantum correlations across causal
boundaries, and non-extensiveness. In fact, we claim that this
entanglement mechanism is not specific to black holes but to any spacetime with a
bifurcating Killing horizon. A d+1 dimensional spacetime that has a d-1
dimensional space-like surface $\Sigma$, on which each point is a fixed point of an
isometry, possesses a bifurcating Killing horizon. This is the null surface generated
by the null geodesics emanating from $\Sigma$. It can be shown that the bifurcating
Killing horizon is invariant under the isometry \cite{KayWald}.

The vacuum state of fields (including fluctuations of the
gravitational field) in a spacetime that has a bifurcating Killing
horizon may be chosen so that the energy-momentum tensor does not
diverge on the horizon. This vacuum state is called the
Hartle-Hawking (HH) state. Examples of spacetimes with a single
bifurcating Killing horizon are de~Sitter space, Kruskal spacetime, the
massive non-rotating BTZ black hole, and Minkowski space (in Rindler coordinates.)

As we shall show, one may view the HH state as an entangled state with respect to
the region of space being observed and an inaccessible region beyond the horizon.
Being resticted to a part of space, an observer will see a mixed rather than pure
state. The entropy of such a mixed state is called entanglement entropy. In
section~\ref{S:duality}, we argue, following the pioneering work of Maldacena
\cite{Maldacena:2001}, that one can interpret the calculation of the number of
states in field theory duals, for a wide class of spacetimes, as a measure of
entanglement entropy.

We  highlight the difference, in principle, between the entropy of a
``microstate gas,'' which conveys information regarding the quantum mechanical
nature of microstates that ``look like" a black hole, and the entanglement entropy,
which is a measure of an observers lack of information regarding the quantum state
of the system in an inaccessible region of spacetime.

Previously, there have been several attempts to relate the entropy and
thermodynamics of the BH to entanglement. In these analyses the entanglement
entropy was interpreted as a one loop correction to the classical Bekenstein
Hawking entropy \cite{SusskindUglum,Callan:1994py,Kabat:1995eq,Frolov:1998vs}. In
contrast, we argue by evaluating the BH entropy using a field theory dual, that the
leading contribution to the entropy is due to entanglement. Our argument depends on
the assumption that the entropy of the field theory dual is equal to the BH
entropy, an equality that is usually a direct consequence of the duality. Our
argument implies that the microstates of the black hole are due to the entangled
nature of the BH vacuum, and are a result of an observers inability to access the
degrees of freedom that are hidden beyond the horizon.

Entanglement of the BH vacuum state was first considered a long time ago by Israel
\cite{Israel}. Here we generalize the path integral method of \cite{KabStr} to
curved spacetimes with non-degenerate, bifurcating Killing horizons. We use our
method to extend the results of \cite{Israel}. Our analysis explicitly reveals the
geometric nature of the vacuum state. This relationship between geometry and
entanglement has been formally discussed in \cite{Jacobson}.

For the case of AdS BH's, the argument that the BH entropy is due to entanglement
follows from the results of Maldacena\cite{Maldacena:2001}. We discuss this case in detail
and relate the divergence of the entanglement entropy to Newton's constant $G_N$.

As we shall discuss in section \ref{S:discussion}, the entanglement interpretation
simplifies some issues regarding area scaling, non-unitary evolution, and the
information paradox. Although it leads to some results that can be obtained using
the ``membrane paradigm" \cite{Thorne:1986iy}, which describes the BH in terms of
a stretched horizon and a thermal atmosphere of the Hawking radiation near the
horizon, it is conceptually different.

In section \ref{S:review}, we review the concept of entanglement and
entanglement entropy in a simplified EPR-like model. In section
\ref{S:spacetime}, we explain and review the entangled nature of the
HH vacuum state in spacetimes with a bifurcating Killing horizon.
In section~\ref{S:duality}, we discuss entanglement for spacetimes
having a field theory dual.  Section \ref{S:AdSBH} deals with the AdS-BH.
We conclude with a discussion of these results.

\section{Review of entanglement entropy}\label{S:review}
In this section we review the concepts of entanglement and entanglement entropy in
quantum mechanics together with the construction of the thermofield double state and its
significance. This section establishes the notation for the rest of the paper.

We start from a simple, well known, example of an EPR type
experiment: suppose we start with two spin half particles in a
singlet state: $| 0,0 \rangle = \frac{1}{\sqrt{2}}\left(| \uparrow
\rangle_1 \otimes | \downarrow \rangle_2 - | \downarrow \rangle_1
\otimes | \uparrow \rangle_2 \right)$. We wish to find the state as
seen by an observer who measures the spin of one of the particles.
Starting with $|0,0 \rangle$, or $\rho=|0,0 \rangle \langle 0,0|$,
and tracing over the spin degrees of freedom of the second particle
we find
\begin{align*}
    \rho_1 &= \text{Trace}_2 \rho \\
           &= \phantom{}_2\langle \uparrow |0,0 \rangle \langle 0,0| \uparrow \rangle_2
            + \phantom{}_2\langle \downarrow |0,0 \rangle \langle 0,0| \downarrow \rangle_2.
\end{align*}
This gives
\[
    \rho_1 =\left(%
        \begin{array}{cc}
            1/2 & 0 \\
            0 & 1/2 \\
        \end{array}%
        \right).
\]
The entropy of the state measured by the second observer will now be
$S_1=- \trace \rho_1 \ln \rho_1 = \ln 2$. Even though we started off
with a pure state with no entropy, an observer restricting her
measurements to part of the available Hilbert space will see a mixed
state with non zero entropy. This entanglement entropy $S_1$ is
intrinsically observer dependant. The microstates being counted are
the, now uncorrelated, spin degrees of freedom. Obviously, this
analysis is symmetric with respect to both Hilbert spaces.

Generally, one defines an entangled state (on a product of two
Hilbert spaces) as a state, that under no local unitary
transformation can be reduced to a direct product of states on the
two Hilbert spaces. Thus $|0,0 \rangle$ is entangled, while $| 1,1
\rangle = | \uparrow \rangle_1 \otimes | \uparrow \rangle_2$ is not.
Starting with a general state $|\psi\rangle =
\sum_{i=1,j=1}^{d,d^{\prime}} A_{ij} |\alpha_i \rangle_1
|\beta_j\rangle_2$, \footnote{In general, the dimensions of the two
subspaces need not be equal. In this paper, we shall use Hilbert
spaces of equal dimension.} one can use Schmidt's decomposition
theorem \cite{Schmidt}  to show that there exists a basis for which
one can write $|\psi\rangle = \delta_{ij} \tilde{A}_{i}
|\tilde{\alpha}_i \rangle_1 |\tilde{\beta}_j \rangle_2$ (the proof
follows by rotating the $\alpha$ and $\beta$ coordinates
independently.)
 Using this basis, we can see that if we take the trace of the initial
state $\rho=|\psi \rangle \langle \psi |$, over one Hilbert space, we obtain
\begin{subequations}
\label{E:rho1and2general}
\begin{align}
    \rho_1 &= \sum_i \tilde{A}^{\star}_i \tilde{A}_i
    | \tilde{\alpha}_i \rangle_1 \, \phantom{}_1\langle \tilde{\alpha}_i | \\
\intertext{or alternately,}
    \rho_2 &= \sum_i \tilde{A}^{\star}_i \tilde{A}_i
    | \tilde{\beta}_i \rangle_2 \, \phantom{}_2\langle \tilde{\beta}_i |.
\end{align}
\end{subequations}
The mixed states associated with $\rho_1$ and $\rho_2$, are the states seen by an observer
restricting her measurements to part of the Hilbert space of states.
Even though originally the entropy of the system was zero, if the
state is entangled, the entropy seen by the observer will be non
zero. Further, as can be seen by eq.~(\ref{E:rho1and2general}), the
entanglement entropy measured by both observers is equal.

When measuring expectation values of an operator $O_r$ that
acts entirely within the Hilbert subspace $\mathcal{H}_r, (r=1,2)$
one observes that \cite{Feynmanstatmech}
\begin{align}
\label{E:trrO_equals_O}
   \langle \psi | O_r  | \psi \rangle &=
 \sum_{i,j}  \tilde{A}_i^{\star}\tilde{A}_j\,
{}_2\!\langle \tilde{\beta}_i|
   {}_1\!\langle \tilde{\alpha}_i |  O_r  |\tilde{\alpha}_j \rangle_1
    | \tilde{\beta}_j \rangle_2\\
    &=\text{Trace}(\rho_r O_r) \qquad \text{(no sum.)}
\end{align}
This implies that instead of considering expectation values of such
operators in an entangled pure state, one may equivalently consider
expectation values of the same operators in a mixed state.

Conversely, instead of considering expectation values of operators
belonging to a Hilbert space $\mathcal{H}$ in a mixed state
$\rho=\sum_i |A_i|^2 |\alpha_i\rangle \langle \alpha_i |$, one can
consider the same expectation values in a state $\psi=\sum_i A_i
|\alpha_i \rangle | a_i \rangle$ defined on an extended Hilbert
space $\mathcal{H} \otimes \mathcal{H}^{\prime}$, where
$| a_i \rangle$ is some arbitrary basis of $\mathcal{H}^{\prime}.$
If the density matrix $\rho$
is to correspond to a thermal state in $\mathcal{H}$, then we should
choose $|A_i|^2=e^{-\beta E_i}/Z$ and
$|\alpha_i\rangle=|E_i\rangle,$ and $E_i$ eigenvalues of the
Hamiltonian $H$ associated with $\mathcal{H}$.
Then we may define the pure state
\begin{equation}
\label{E:thermointermediate}
    |\psi\rangle =\frac{1}{\sqrt{Z}}
    \sum_i e^{-\beta/2 E_i} |E_i \rangle |a_i\rangle.
\end{equation}
The factor of two in the temperature appears since the weights of the thermal state
are the squares of the amplitude of the entangled state.

If we further wish that the state defined in equation
(\ref{E:thermointermediate}) be invariant under time translations,
then may take the Hilbert space $\mathcal{H}^{\prime}$ to be
a duplicate of $\mathcal{H},$ but take the global
Hamiltonian to be $H_{global}=H-H'.$
The state obtained in this way is
\begin{equation}
\label{E:thermofielddouble}
    |\psi\rangle =
    \frac{1}{\sqrt{Z}} \sum_i e^{-\beta/2 E_i} |E_i \rangle |E_i\rangle^\prime,
\end{equation}
called a thermofield double \cite{TFD} (TFD). This procedure is
also known as purification in the quantum information literature
\footnote{We thank D. Terno for pointing this out}. We shall use
thermofield doubles extensively in what follows.

Note that eq.~(\ref{E:trrO_equals_O}) is valid
regardless of the thermal nature of $\rho$ and that one may create
an entangled state to describe any $\rho$ (and vice versa)  using
the above method. Thus, the thermal field double appears to be artificial.  Remarkably,
in some spacetimes, it turns out to be the natural physical state, as we shall now discuss.

\section{Entanglement in Spacetimes}\label{S:spacetime}

\subsection{Non-degenerate Single Horizon}\label{S:nondegenerate}
Shortly after the discovery that black holes radiate, it was shown
\cite{Israel} that the vacuum of  spacetimes with a single
bifurcating Killing horizon is a TFD state, with respect to the left
and right wedges of the spacetime. In this section we elaborate on
these results, using a path integral method that was originally
introduced in \cite{KabStr}, showing that the natural structure of
the Hilbert space of fields in a curved background with a
bifurcating Killing horizon is a product of two Hilbert spaces, one
for each wedge of the spacetime, and that the Hartle-Hawking vacuum state is the
TFD state. This method offers a geometric interpretation of the
doubling of the Hilbert spaces and allows an extension of the result
to rotating and charged black holes.

We start from a metric of the form:
\begin{equation}
\label{E:bifkilling}
    ds^2=-f(r)d\tau^2+dr^2/f(r)+q(r)dx_{\bot}^2,
\end{equation}
with the radial coordinate restricted to the region $r \ge r_s$ ($f(r_s)=0$.) The
coordinate transformation
\begin{align}
\label{E:coord_trans}
    x&=\sqrt{g(r)} \cosh (\tau f^{\prime}(r_s)/2)\\
    t&=\sqrt{g(r)} \sinh (\tau f^{\prime}(r_s)/2),
\intertext{with}
    g&=\frac{4 \xi_0 }{f^{\prime\,2}(r_s)} e^{f^{\prime}(r_s) \int^r \frac{1}{f} dr}
\end{align}
gives us the metric
\begin{equation}
\label{E:bifkilling_kruskal}
    ds^2=h(r)(-dt^2+dx^2)+q(r)dx_{\bot}^2.
\end{equation}
where $h(r)= \frac{1}{\xi_0} f(r) e^{- f^{\prime}(r_s)\int^r
\frac{dr'}{f(r')}}$ and $\xi_0$ is an integration constant. We can
see explicitly that $h(r_s)\neq 0$ if $f(r)$ has a simple zero at
$r_s$, and (assuming that the metric is analytic in $x$ and $t$)
 that the coordinate patch now covers all of spacetime
(except perhaps for a real singularity.)

In Lorentzian signature, the metric in equation (\ref{E:bifkilling_kruskal}) covers
all four wedges of the spacetime, while the metric in (\ref{E:bifkilling}) covers
one wedge. Therefore, we shall call the coordinates in (\ref{E:bifkilling_kruskal})
global coordinates, and the coordinates in (\ref{E:bifkilling}) wedge coordinates.
In Euclidean signature the wedge coordinates cover all of Euclidean spacetime.

 Consider now a (scalar) field in the background of (\ref{E:bifkilling_kruskal}).
Under the preceding assumptions, one may define \cite{Jacobson}
 the Hartle-Hawking vacuum state  $| 0 \rangle$
by switching to a Euclidean signature,  in which the wave functional
of the vacuum is
\[
      \bra{0} \psi(\vec{x}) \rangle =
      \int_{\varphi(\vec{x},0)=\psi(\vec{x})}
            \exp\left[-\int_0^{\infty}\ldots\int\mathcal{L} d^dx d\tau\right]
      D\varphi,
\]
where the integral is over all fields evaluated at $t > 0$ that
satisfy the boundary conditions $\varphi(\vec{x},0)=\psi(\vec{x})$.

Next we define the density matrix
\[
    \rho_{R}=\trace_{L}\ket{0}\bra{0},
\]
where $R,L$ correspond to the right and left wedges of spacetime.
We wish to show that for the above spacetime $\rho_{R} = e^{-\beta
H_R}$, where $H_R$ is the wedge Hamiltonian. We start by looking for
a Euclidean path integral representation of the elements of
$\rho_{R}$, that is, of $\bra{\psi^{\prime}_{R}} \rho_{R}
\ket{\psi^{\prime\prime}_{R}}$. In the specific case of interest,
\[
    \langle 0 |\psi_{R}\psi_{L}\rangle=\int_
        {\varphi(\vec{x},0)=
            \begin{cases}
                    {\scriptstyle x > 0, t=0} & {\scriptstyle \psi_{R}(\vec{x})} \\
                    {\scriptstyle x < 0, t=0} &  {\scriptstyle \psi_{L}(\vec{x})} \\
            \end{cases}}
        \exp\left[-\int_0^{\infty}\ldots\int\mathcal{L} d^dx d\tau\right]
        D\varphi.
\]
It follows that for $\ket{\psi} = \ket{\psi_{R} \psi_{L}}$, we have
\begin{equation}
\label{E:rho_global}
 \bra{\psi^{\prime} } \rho
\ket{\psi^{\prime\prime}}
    =\int_
        {\varphi(\vec{x},0)=
            \begin{cases}
                    {\scriptstyle x > 0, t=0^{+},} & {\scriptstyle \psi_{R}^{\prime}(\vec{x})} \\
                    {\scriptstyle x > 0, t=0^{-},} & {\scriptstyle \psi_{R}^{\prime\prime}(\vec{x})}\\
                    {\scriptstyle x < 0, t=0^{+},} & {\scriptstyle \psi_{L}^{\prime}(\vec{x})} \\
                    {\scriptstyle x < 0, t=0^{-},} & {\scriptstyle \psi_{L}^{\prime\prime}(\vec{x})}
            \end{cases}}
        \exp\left[-\int_{-\infty}^{\infty}\ldots\int\mathcal{L} d^dx d\tau\right]
        D\varphi.
\end{equation}

Working in a Euclidean signature, we may use the Euclidean version
of the wedge coordinates (\ref{E:bifkilling}) to cover the entire
spacetime instead of the global coordinates. This would be
equivalent to changing from flat coordinates in the $x-t$ plane, to
a `stretched' (due to the $g(r)$ term in the transformation
(\ref{E:coord_trans})) polar $r - \tau$ coordinate system. This will
give us
\begin{align*}
    x>0 ,\, t&=0^{\pm} \to r>r_s ,\, \tau=0^{\pm}\\
    x<0 ,\, t&=0^{-} \to r>r_s ,\, \tau=-\beta/2\\
    x<0 ,\, t&=0^{+} \to r>r_s ,\, \tau=\beta/2
\end{align*}
so that in these coordinates equation (\ref{E:rho_global}) changes
to
\begin{equation}
\label{E:rho_wedge}
    \bra{\psi^{\prime} } \rho
\ket{\psi^{\prime\prime}}
    =\int_
        {\varphi(\vec{x},0)=
            \begin{cases}
                    {\scriptstyle r>r_s, \tau=0^{+},} & {\scriptstyle \psi_{R}^{\prime}(\vec{x})} \\
                    {\scriptstyle r>r_s, \tau=0^{-},} & {\scriptstyle \psi_{R}^{\prime\prime}(\vec{x})}\\
                    {\scriptstyle r>r_s, \tau=\beta/2,} & {\scriptstyle \psi_{L}^{\prime}(\vec{x})} \\
                    {\scriptstyle r>r_s, \tau=-\beta/2,} & {\scriptstyle \psi_{L}^{\prime\prime}(\vec{x})}
            \end{cases}}
        \exp\left[-\oint \ldots\int\mathcal{L} d^dx d\tau\right]
        D\varphi.
\end{equation}
The integral over the $\tau$ coordinate is now on a circular path from
$\tau=-\beta/2$ to $\tau=\beta/2$. Geometrically, this path integral is a product
of two path integrals each over half a torus. One radial coordinate of the torus is
the time coordinate, while the other specifies the spatial coordinates. The
boundary conditions are such that the two half tori can not be glued together.

\begin{figure}
\begin{center}
\includegraphics{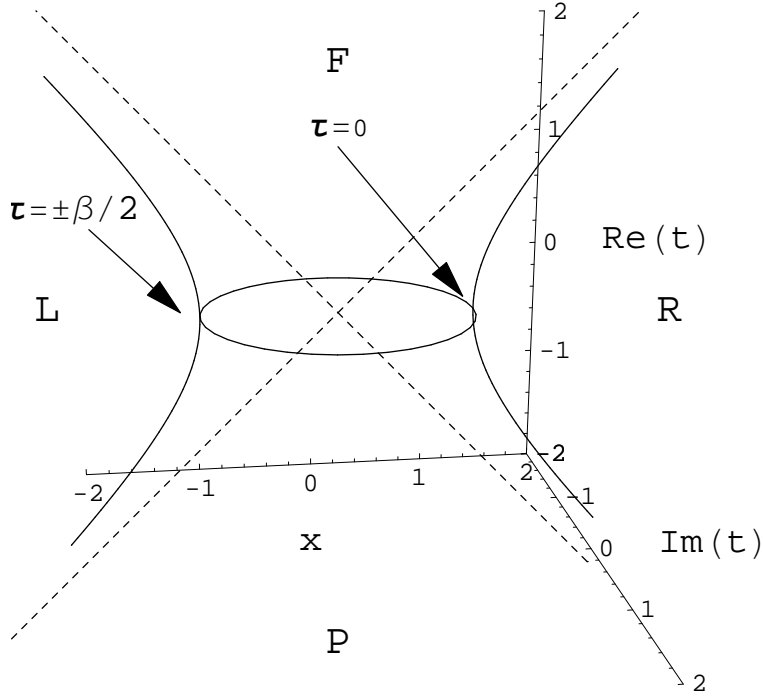}
\caption{Diagram of a line of constant radial coordinate $r$ in complex time. The
circle is a line of constant $r$ in imaginary time where $\tau$ is an angle, while
the hyperbolas are lines of constant $r$ in real time. The dashed lines represent,
in real time, the bifurcation surface that splits the spacetime into the four
wedges R, L, F, P.} \label{F:geometric}
\end{center}

\end{figure}

Now, the elements of $\rho_{R}$ are given by
\begin{align*}
    \rho_{R} &= \sum_{\psi_{L}} \bra{\psi_{L}} \rho \ket{\psi_{L}}\\
              &= \sum \langle {\psi_{L}} \ket{\psi^{\prime}}  \bra{\psi^{\prime}}
                    \rho \ket{\psi^{\prime\prime}}  \langle {\psi^{\prime\prime}}
                 \ket{\psi_{L}}
\end{align*}
so
\begin{equation}
\label{E:rhoin}
    \bra{\psi_R^{\prime}}\rho_{R}\ket{\psi_R^{\prime\prime}}
    =\int_
        {\varphi(\vec{x},0)=
            \begin{cases}
                    {\scriptstyle r>r_s, \tau=0^{+},} & {\scriptstyle \psi_{R}^{\prime}(\vec{x})} \\
                    {\scriptstyle r>r_s, \tau=0^{-},} & {\scriptstyle \psi_{R}^{\prime\prime}(\vec{x})}\\
            \end{cases}}
        \exp\left[-\oint \ldots\int\mathcal{L} d^dx d\tau\right]
        D\varphi.
\end{equation}
Now that the trace has been performed the two half tori have been glued at
$\tau=\pm\beta/2$.

We shall show below that all the matrix elements of $\rho_R$ are equal the matrix
elements of $e^{-\beta H_R}$: $\bra{\psi^{\prime} } \rho
\ket{\psi^{\prime\prime}}=\bra{\psi^{\prime} } e^{-\beta H_R}
\ket{\psi^{\prime\prime}}$.

In the $r - \tau$ coordinates $H_R$, the generator of time translations, is time
independent. So, we may use the standard time-slicing technique for expressing
$e^{-\beta H_R}$ as a path integral in imaginary time from $\tau=0$ to
$\tau=-\imath \beta_0$.
\[
    \langle\psi_{in}^{\prime}|e^{-\beta H_R}|\psi_{in}^{\prime\prime}\rangle
        =\int_{\substack{\varphi(x,0)=\psi_{in}^{\prime}(\vec{x}) \\
            \varphi(x,-i\beta_0)=\psi_{in}^{\prime\prime}(\vec{x})}}
        \exp\left[
            \imath\int_0^{\imath\beta_0} \left(
                \int \pi \frac{d\varphi}{dt} d^dx - H_R \right) dt
            \right]
        D\varphi D\pi.
\]
Changing to Euclidean time:
\begin{equation}
\label{E:e_bheff_generalform}
    \langle e^{-\beta H} \rangle
    =\int_{\substack{\varphi(x,0)=\psi_{in}^{\prime}(\vec{x}) \\
            \varphi(x,\beta_0)=\psi_{in}^{\prime\prime}(\vec{x})}}
        \exp\left[
            \int_0^{\beta_0} \left(
                \int \imath \pi \dot{\varphi} d^dx - H \right) (d\tau)
            \right]
        D\varphi D\pi.
\end{equation}

Since $H$ is the generator of time translations, it is the Legendre
transform of a Lagrangian. A Lagrangian density for a scalar field
in a gravitational background with metric $g_{\mu\nu}$ is given by $
    \mathcal{L}=-\frac{1}{2}g^{\mu\nu}
        \partial_\mu\varphi \partial_\nu\varphi-V(\varphi)
$. The Hamiltonian is $H=\int\mathcal{H}\sqrt{g}d^dx\,dt$, where $
    \mathcal{H}=\frac{1}{2}
    \left(\left(\frac{\pi}{g^{00}}\right)^2+h^{ij}\partial_i\varphi\partial_j\varphi\right)
    +V(\varphi),
$ $h^{ij}$ are the space-space components of the metric ($h^{ij}=g^{ij}$), and
$\pi=g^{00}\dot{\varphi}$.

Performing the Gaussian integral and using
$\Omega=\frac{1}{\sqrt{g_{00}^E}}$, (where $g_{00}^{E}=-g_{00}$),
one obtains \cite{deAlwis1},
\begin{equation}
\label{E:rhothermal}
    \langle\psi_{in}^{\prime}|e^{-\beta H}|\psi_{in}^{\prime\prime}\rangle=
    \int_{\substack{\varphi(\xi,0)=\psi_{R}^{\prime}(\xi) \\
            \varphi(\xi,\beta_0)=\psi_{R}^{\prime\prime}(\xi)}}
        \exp\left[
            -\int_0^{\beta_0} \int
                \mathcal{L}_{E}
            d^d \xi d\eta
            \right]
        \left|\Omega\right|\,\left|g\right|^{\frac{1}{4}}
        D\varphi.
\end{equation}
Obviously, equations (\ref{E:rhoin}), and (\ref{E:rhothermal}) are equal up to the
Liouville term.

We found that the matrix elements of $\text{Trace}_R | 0 \rangle \langle 0 |$ are
equal to those of $ e^{-\beta H_R}$. From symmetry arguments, the same applies to
$\text{Trace}_L | 0 \rangle \langle 0 |$. It follows that the Hartle-Hawking state
$| 0 \rangle$ is a TFD state.

Now that we know what the vacuum state looks like, we can find the
entanglement entropy that is associated with the inaccessibility to
the region of space beyond the horizon. This is given by $S_R =
\text{Tr}(\rho_R \log \rho_R)$. Since $\rho_R = e^{-\beta H_R}$,
this is simply the standard thermodynamical entropy associated with
the fields in the wedge of spacetime, and was evaluated in
\cite{deAlwis2}. In the high temperature limit, the
entropy of a single scalar field in a curved background in $d+1$
dimensions is given by $S=\frac{(d+1)\,T^d V_{d}}{\pi^{(d+1)/2}}
\Gamma\left( \frac{d+1}{2} \right) \zeta(d+1),$ where $V_d$ is the
volume of space in the optical metric (obtained by a conformal
transformation on the metric that leaves $g_{00}=1$.) This entropy
is formally infinite due to the divergence of the optical volume
from the region near the horizon. Using (\ref{E:bifkilling}) and
setting $T = \frac{f^{\prime}(r_s)}{4 \pi}$ as the Hawking
temperature, one finds
\begin{equation}
\label{E:entanglemententropy_scalar}
    S =\frac{q(r_s)^{\frac{d-1}{2}} (d+1) \Gamma\left(\frac{d+1}{2}\right)
    \zeta(d+1)}{2^{2d-1} \pi^{\frac{3d+1}{2}}(d-1)} \delta^{-(d-1)}
    \int d\vec{x}_{\bot}.
\end{equation}

The short distance cutoff away from the horizon, $\delta$, is in
units of proper distance. Had there been several fields that
contribute to the entropy, and not only one scalar field, we would
have found
\begin{equation}
\label{E;entanglemententropy}
    S=\widetilde{C}\,N_{\rm eff}\,\delta^{-(d-1)} A_{\rm hor}.
\end{equation}
where $N_{\rm eff}$ takes into account the contributions of the various fields and
$\widetilde{C}$ is a numerical factor. We see that the entanglement entropy is
proportional to the surface of the horizon, and that it diverges as we allow the
volume of optical space to enclose the regions of space that are infinitesimally
close to the horizon surface. This ultra-violet (UV) divergence was first pointed
out in \cite{'tHooft:1985}. We shall discuss this further in the context of the
AdS/CFT correspondence in section \ref{S:AdSBH}.

\subsection{Non-degenerate multiple killing horizons}
For spacetimes with multiple Killing horizons, such as the charged or rotating
black holes there are some slight subtleties: we consider a spacetime with the
metric
\begin{equation}
\label{E:mulkilling}
    ds^2=-f(r)d\tau^2+dr^2/f(r)+q(r)dx_{\bot}^2,
\end{equation}
where now there are two horizons at $f(r_{\pm})=0$, where
$f^{\prime}(r_{\pm}) \neq 0$. An example would be the
Reissner-Nordstrom solution for a charged black hole. If the inner
horizon ($r_-$) is a cauchy surface for the observer at $r>r_-$,
then one may consider only those four wedges on which the initial
value problem for the field equation is well defined. In this case
one can define a vacuum state, and follow through the procedure
described above to show that the locally defined vacuum is the TFD
of the left and right wedges for the observer at $r_-<r<\infty$. One
should note however, that such a vacuum state is not well defined in
all of spacetime \cite{KayWald,Jerusalemgroup}.

We proceed as in the previous section: the expression for $\rho_R$
will be unchanged (eq.~(\ref{E:rhoin})). Though here the energy
density: $E = T^{0 \nu} K_{\nu}$ (where $K_{\nu}$ is the Killing
vector for time translations and $T^{\mu\nu}$ is the energy momentum
tensor) is not the same as the Hamiltonian density. For a scalar
field we find $\mathcal{H} = E + ( \imath e A_0 \pi \varphi +
c.c.)$, with $\pi$ the canonical momentum. For an electrically
charged black hole we may write this last term as $e A_0 \sqrt{g}
j^{0}$, where $j^{\mu}=\imath(\pi \phi-\pi^{\dagger}\phi^{\dagger})$
is the scalar field charged current density. On comparing equations
(\ref{E:rhoin}) and (\ref{E:rhothermal}) we find
\begin{equation}
    \rho_R = e^{-\beta E - \beta \int e A_0 j^0 \sqrt{g} d^dx}.
\end{equation}
where the last term may be interpreted as a chemical potential.

We are unaware of a canonical form for the metric of rotating black holes. However,
for all rotating black holes that we are aware of, it is possible to find a
coordinate transformation similar to those in eq.~(\ref{E:coord_trans}) that extend
the wedge geometry to the global one. An explicit example of such a transformation
in the BTZ BH can be found in \cite{BTZ}. So, following the same considerations as
before we expect that $\rho_R$ for the case of rotating BH's is also given by
eq.(\ref{E:rhoin}).

As opposed to the charged case, for rotating BH's $E=\mathcal{H}$.
However, in the derivation of eq.~(\ref{E:rhothermal}) a
complication arises: the time and angular directions are not
orthogonal. This requires the extension of the standard path
integral construction, since in the standard derivation one usually
assumes that $H$ will generate translations in the time direction
only. If the Hamiltonian generates spatial (or in this case,
angular) translations as well as time translations we find that, for
a scalar field, $\langle \psi(x^{\prime},t^{\prime}) | \psi(x, t
\rangle) = \int D\phi D\pi \exp \left({\imath \int \left( \int
\left(\pi \dot{\phi} + N^i \partial_i \phi \pi \right) d^d x - H
\right)dt}\right) $ where $N^i = \frac{g^{0i}}{|g^{00}|}$ is the
shift vector. This will then give us:
\begin{equation}
    \rho_R = e^{-\beta E + \beta\int N^i \Pi_i d^dx}
\end{equation}
where $\Pi_i$ is the momentum in the $i$ direction. The last term
here may also be interpreted as a chemical potential. For a metric
describing uniform rotation with angular velocity $\Omega$, one has
$N^\theta=\Omega$, so $N^i \Pi_i =\Omega L_\theta$, where
$L_{\theta}$ is the angular momentum in the $\theta$ direction.

To summarize, we find that the non-extremal charged and rotating black holes have a
vacuum structure similar to the non-rotating non-charged black hole. The partition
function that is generated by tracing over the degrees of freedom in the adjacent
spacetime wedge has a chemical potential, as expected on general grounds. Hence our
results apply to rotating and charged BHs as well.

\subsection{Extremal black holes}

We define extremal black holes as spacetimes of the form
(\ref{E:bifkilling}) but here $f(r_s)=0$ and $f^{\prime}(r_s)=0$.
This may be viewed as a spacetime with multiple Killing horizons
(\ref{E:mulkilling}) where the limit $r_+ \to r_-$ has been taken.
As the horizons approach each other, the throat (Einstein bridge)
connecting the two left and right wedges seen by the $r>r_-$
observer becomes longer and longer, until it becomes infinitely long
when $r_+ = r_-$ (see for example \cite{Townsend}.) From this point
of view one should consider two disconnected extremal black holes.
The equivalent of the HH state in this case is defined as the TFD
state of the extremal black hole spacetime. Here, as opposed to the
non-extremal case, the two components of the doubled Hilbert space
are geometrically disconnected. We shall not discuss this case
further.

\section{Entanglement in field theory duals}\label{S:duality}

Black hole entropy has been calculated in string theory using D-branes (originally,
in ref.~\cite{StromVafa}.) These provide a gauge theory dual to string theories in
a curved background. The AdS/CFT is the best understood duality of this kind, and
black hole entropy has been calculated in this setup as well \cite{Peetetal}. In
the AdS/CFT correspondence, the black hole entropy is related to the thermal
entropy of the CFT at an appropriate temperature.

We  relate such entropy calculations to entanglement entropy. To this
end, we argue that the field theory dual of the black hole (global) vacuum is a TFD
state (as originally suggested in \cite{Maldacena:2001} in the context of the
AdS/CFT correspondence.) This implies that the entropy of a thermal state in the FT
may be interpreted as entanglement entropy and, as a  consequence, that the black
hole entropy may be interpreted as entanglement entropy. Thermal entropy may always
be interpreted as coming from a fictitious extra Hilbert space. This was the
original motivation for introducing a TFD \cite{TFD}. However, for the case of
black holes, this extra Hilbert space has geometric meaning---it is a reminder of
the degrees of freedom beyond the horizon.

We start with a stringy description of a gravitational system in a background
spacetime $X$, with a partition function $Z_{string}(X)$, and a field theory $M$
with partition function $Z_{FT}(M)$. That the theories are dual, implies that
$Z_{string}(X) = Z_{FT}(M)$. In the low energy ($\alpha^{\prime} \to 0$), tree
level ($g_s \to 0$), limit, the string theory partition function is approximately
equal to the supergravity partition function $Z_{string} \approx e^{-I_{GRAVITY}}$.
Under the AdS/CFT correspondence, this is dual to the strong coupling limit,
$\lambda_{YM} \to \infty,$ and $N \to \infty$ of the CFT. In what follows we shall assume that duality
implies an isomorphism of the string and field theory Hilbert spaces.  We assume that the
isomorphism of Hilbert spaces should hold for any value of $\alpha^{\prime}$ and
$g_s$, including the low energy limit $\alpha^{\prime} \to 0$, $g_s \to 0$.

Suppose that $X$ is a spacetime with a bifurcating Killing horizon,
and $X_W$ is a wedge of this spacetime. Correspondingly, let $M$ be the
field theory dual of the string theory on $X,$ and $M_W$ be the field
theory dual of the string theory on $X_W.$  Let us denote the Hilbert
space of small fluctuations of fields (including the gravitational
field) around $X$ by $\cal{H} = \cal{H}_W \otimes \cal{H}_W$.
In the low energy limit, tracing over the degrees of freedom in the left wedge
will result in a thermal state on the right wedge and vice versa. The partition
function on the right wedge in this limit is $e^{-I_{GRAVITY}(X_W)} = Z(X_W)$,
where the density matrix is given by $\rho_R = \trace_{L} | 0 \rangle \langle 0 |$.
So, one can go from $e^{-I(X)}$ to $e^{-I(X_W)}$ by tracing over states in the
left wedge. We shall denote this tracing procedure as
\begin{equation}
\label{E:thermal3}
    e^{-I(X)} \longrightarrow_{\hspace{-15pt}_{Tr}\hspace{15pt}}
    e^{-I(X_W)}.
\end{equation}
As a result of the tracing procedure, the boundary
conditions are different for the associated path integrals. The first has
boundary conditions on the boundary of the extended space $\partial
X,$ while the second on the boundary of a wedge $\partial X_W$. The
symbol $\longrightarrow_{\hspace{-15pt}_{Tr}\hspace{15pt}}$
expresses the fact that to go from a partition function on the
extended space $X$, $e^{-I (X)}$, to the partition function on a
wedge $X_W$, $e^{-I (X_W)}$, we may take the state defining the
first partition function, trace over half of its degrees of freedom,
and evaluate the second partition function in a state defined by the
resulting density matrix.

Let us consider the low energy limit of the dual field theory. The partition
function is $Z_{FT}(M_{W})$, and the density matrix is given by
\begin{equation}
\label{E:def_of_rhoFT}
    \widetilde{\rho}_{W}=\frac{1}{Z_{FT}}
    \sum_i e^{-\beta \widetilde E_i}| \widetilde E_i \rangle \langle \widetilde E_i |.
\end{equation}
In this low energy limit, the Hilbert space on the extended
spacetime $\cal{H}$ is isomorphic to the Hilbert space in the dual
field theory $\widetilde{\cal{H}}$, and $\cal{H}_W$ is isomorphic to
$\widetilde{\mathcal{H}}_W$ so we must have
$\widetilde{\mathcal{H}}=\widetilde{\mathcal{H}}_W \otimes
\widetilde{\mathcal{H}}_W$.

We observe that
\begin{equation}
\label{E:traceft}
    Z_{FT}(M) =
        e^{-I (X)} \longrightarrow_{\hspace{-15pt}_{Tr}\hspace{15pt}}
         e^{-I (X_W)}
    =Z_{FT}(M_W).
\end{equation}
Therefore, $\widetilde{\rho}_W$ is obtained from $\widetilde{\rho}$ (the dual of
$\rho=|0\rangle \langle 0 |$) by an operation that is dual to tracing over the
degrees of freedom in the left wedge of the vacuum state $| 0 \rangle \langle 0 |$.
The dual of tracing over the degrees of freedom in $\mathcal{H}_W$ should be
tracing over the degrees of freedom in $\widetilde{\mathcal{H}}_W$. Let the dual of
$| 0 \rangle$ be $| \widetilde{0} \rangle$. Since the Hilbert space $\mathcal{H}_W$
is isomorphic to $\widetilde{\mathcal{H}}_W$ ($\mathcal{H}_W \cong
\widetilde{\mathcal{H}}_W$), it follows that for $\{ | \psi_{L,i} \rangle \}$ a
complete set in $\mathcal{H}_W$ and $\{ | \widetilde{\psi}_{L,i} \rangle \}$ its
dual in $\widetilde{\mathcal{H}}_W$ that
\begin{equation}
\label{E:traceft1}
    \sum_i  \langle {\psi_L}_i  | 0 \rangle \langle 0 | {\psi_L}_i \rangle
    \cong
    \sum_i  \langle \widetilde{\psi}_L { }_{i}  | \widetilde{0} \rangle
        \langle \widetilde{0} | \widetilde{\psi}_L {  }_{i} \rangle,
\end{equation}
so
\begin{equation}
\rho_W=\label{E:traceft2} \text{Tr}_L  | 0 \rangle \langle 0 | \cong
\text{Tr}_L | \widetilde{0} \rangle \langle \widetilde{0} |.
\end{equation}
eq.~(\ref{E:traceft}) then reduces to $Z_{FT}(M)
\longrightarrow_{\hspace{-15pt}_{Tr}\hspace{15pt}} Z_{FT}(M_W)$. Therefore
$\widetilde{\rho}_{W} = \text{Tr}_L | \widetilde{0} \rangle \langle \widetilde{0} |
$. Since tracing over one Hilbert space $\widetilde{\mathcal{H}}_W$ of the state
$|\widetilde{0} \rangle$ results in a thermal state in the other, it follows that
the $|\widetilde{0} \rangle$ state is the TFD of the FT on
$\widetilde{\mathcal{H}}_W \otimes \widetilde{\mathcal{H}}_W$.

We have now shown that the vacuum of the field theory $M$ that is dual to string
theory on the extended spacetime $X$ is (in the low energy limit $\alpha^{\prime}
\to 0$, $g_s \to 0$) a TFD state. We have done this by appealing to the isomorphism
of Hilbert spaces in the two descriptions, and to the special, entangled, nature of
the HH vacuum.
Let us now use this information to interpret black hole entropy as coming from
entanglement. Suppose $X_W$ is a spacetime whose entropy is $S(X_W)=\frac{A}{4
G_N}$. Now consider the field theory dual $M_W$. Its (thermal) entropy should also
be $S(M_W)=\frac{A}{4G_N}$. The entropy $S(M_W)$, as we have just shown, can be
interpreted as due to entanglement in the initial TFD state $|\widetilde{0}\rangle$
of $M$. This implies that $S(X_W)$ is also due to entanglement of the $|0\rangle$
state.

To summarize, we find a chain of equalities. First,  the
entanglement entropy on the gravity side is equal to the
entanglement entropy on the FT side $S_{\rm entanglement,\
gravity}=S_{\rm entanglement,\ FT}$. Then, we know that the
entanglement entropy on the FT side is equal to the thermal entropy
of one FT at temperature $T>0$: $S_{\rm entanglement,\
FT}=S_{T>0,FT}$. Furthermore, we know that the thermal entropy of
the FT is equal to the BH entropy $S_{T>0,FT} =S_{BH}$. The final
result is that $S_{\rm entanglement,\ gravity}=S_{BH}$.

Note that the entanglement entropy, defined at $G_N \to 0$, diverges
as does the Bekenstein-Hawking entropy in this limit. To define and
compare them at finite $G_N$ one may use the dual theory, for which
all quantities are well defined, as we have just done.

Referring to our earlier discussion surrounding eq.(\ref{E;entanglemententropy}),
 consistency therefore requires a relationship between $G_N$, the number of light
fields, and the short distance cutoff scale. There are several indications that
such a relationship should exist. The first appearance of such a relation was in
\cite{'tHooft:1985}. An indication of a similar relation using entropy bounds
appeared in \cite{Shortest}. A different relationship between the maximal curvature
of the universe, the number of light fields and the Planck scale, appears in
\cite{beb1,gsl,gslstring,Brustein:2001di}. More recently, such connections have
been found in braneworld models \cite{HawMalStr,Branepuzzle}.

\section{AdS black holes - an example}
\label{S:AdSBH}%

The entropy of the five dimensional AdS black hole in the wedge coordinates has
been calculated either using the Hawking formula: $S_{BH}=\frac{A}{4 G}$ or
perturbatively in the CFT dual \cite{Peetetal}, which gives $S_{CFT}=4/3 S_{BH}$.
The different factor ($4/3$) arises because the Bekenstein-Hawking entropy and the CFT entropy
have been calculated at two different points in the parameter space of the theory.
Similar calculations for BH's in other dimensions were also performed
\cite{AdSCFTreview}.

Maldacena \cite{Maldacena:2001} has shown that the dual of the product theory
CFT$\otimes$CFT is approximately an AdS black hole in global coordinates. The
approximation is related to contributions coming from sub-leading topologies. These
were conjectured to have an important effect when considering Poincare recurrences,
and were discussed in detail in \cite{BarbonRab,PRK}. Here we will consider very
massive AdS black holes for which the boundary is $ \mathbb{R}^{n-1} \times S^1 $
rather than $ S^{n-1} \times S^1 $ \cite{Witten9803}. In this case there are no
known contributions from other topologies to the partition function. In what
follows, we shall therefore ignore contributions from sub-leading topologies.

A direct computation of the entanglement entropy of the AdS black hole in the low
energy approximation ($\alpha^{\prime} \to 0$) yields a divergent result (see
eq.(\ref{E;entanglemententropy}), and below) due to contributions from distances
that are very close to the horizon.

Alternately, one may calculate the entanglement entropy of the AdS
black hole via its CFT dual. In order to do this one should start
off with the dual of the HH state $| \widetilde{0} \rangle$, and
trace over the Hilbert space that is dual to one of the wedges
$\widetilde{\cal{H}}_W$. The entanglement entropy seen by an
observer, is the entropy associated with the state defined by
$\widetilde{\rho}_W = \trace_L \ket{\widetilde{0}}
\bra{\widetilde{0}}$. Since the state $| \widetilde{0} \rangle$ is
a TFD then $\widetilde{\rho}_W=\frac{1}{Z}e^{-\beta
\widetilde{H}_W}$, so that the entanglement entropy is simply the
entropy in a canonical ensemble of the CFT. By dimensional
analysis this entropy is equal to $f(\lambda_{YM}) N^2 V T^3$. The
function $f(\lambda_{YM})$ has been evaluated in \cite{Peetetal},
and agrees with the Hawking formula, up to numerical coefficients,
which are related to the evaluation of $f(\lambda_{YM})$ in different
points of parameter space. Therefore, dualizing back to the
AdS black hole, we find that the entanglement entropy agrees with
the Bekenstein-Hawking formula (up to the ambiguities discussed
above.)

To make the comparison in detail, let us consider the case of the 5D AdS BH. Recall
that for large black holes in 5D AdS the Einstein frame metric is approximately
give by
\begin{equation}
ds^2= -\left(\frac{r^2}{R^2}-\frac{b^4 R^2}{r^2}\right) dt^2
+\left(\frac{r^2}{R^2}-\frac{b^4 R^2}{r^2}\right)^{-1} dr^2+ \frac{r^2}{R^2}\left(
dx_1^2 + dx_2^2+ dx_3^2\right).
 \label{emetric5}
\end{equation}
The constant $b\gg 1$ is related to the 5D Newton's constant $G_5$, to the mass of
the BH $M_{BH}$,  to the AdS scale $R$, $ b^4=\frac{8 G_5}{3 \pi}
\frac{M_{BH}}{R^2}$, and to the BH temperature $ b\simeq \pi R T_{BH}$. The BH
horizon is at $r_s\simeq b R$. The entropy of the BH is given by
\begin{equation}
S_{BH}=\frac{A_{\rm hor}}{4G_5}=C  \frac{ R^5\, A_{\rm hor}}{l_s^8
g_s^2},
 \label{bhentropy5}
\end{equation}
where $C$ is a convention dependent numerical factor that determines
the relationship between the string length and Newton's constant in
10D and $l_s$ is the string scale. The area of the horizon is given
by $A_{\rm hor}=\frac{r_s^3}{R^3} \int dx_1 dx_2 dx_3$.

As we have discussed, the entanglement entropy $S_{\rm entanglement,\ gravity}$ in
eq.(\ref{E;entanglemententropy}) is equal to the Bekenstein-Hawking entropy
$S_{BH}$ in eq.(\ref{bhentropy5}). This gives the following relation
\begin{equation}
\label{E:Grelation}
  \widetilde{C} \, N_{\rm eff}\, \delta^{-3} = C \frac{R^5}{l_s^8
  g_s^2}.
\end{equation}

Equation (\ref{E:Grelation}) may be justified as follows. Recall that we are using
the low energy approximation, and that the 5D metric is only a part of the full 10D
metric that describes a BH in $AdS_5\times S^5$. In order for the low energy
approximation to be valid, we need that massive string modes do not contribute to
the action. This gives, in the Einstein frame, $\delta \sim g_s^{1/4} l_s\,=l_p$
(where $l_p$ is the Planck length.) This cutoff scale is much higher the
Kaluza-Klein scale. Consequently we need to include in $N_{\rm eff}$ the full tower
of Kaluza-Klein modes on the five sphere. For each 10D massless field  we get
$N^{(5)}_{\rm eff}\sim \frac{R^5}{\delta^5}$  fields in 5D, hence $N_{\rm eff}= K
\frac{R^5}{\delta^5}$ ($K$ is a numerical factor). Substituting the expressions for
$N_{\rm eff}$ and $\delta$ into $N_{\rm eff}\, \delta^{-3}$ we obtain
eq.(\ref{E:Grelation}). This analysis generalizes to higher dimensional black
holes.

We have compared the bulk theory provided with a short
distance cutoff to a theory with finite $N$ and $\lambda_{YM}$ on
the boundary. The short distance cutoff and finite N were used to
regulate the entropy. Since currently we do not know of a precise
relation between the two regularization prescriptions, we can only
obtain a relationship between the entanglement entropy and the
Bekenstein-Hawking entropy up to an undetermined numerical
factor.

\section{Discussion}\label{S:discussion}
Due to the structure of the Hilbert space of states of fields in a
spacetime background that admits a bifurcating Killing horizon, an observer in
part of the spacetime will observe a density matrix reflecting her
lack of information regarding the region beyond the horizon.
For thermal states, the associated entropy is called entanglement entropy.  From this
point of view, for eternal BH's, the entanglement entropy is the entire black hole entropy, in contrast to
previous interpretations where entanglement entropy was considered to be a quantum
correction to the black hole entropy
\cite{SusskindUglum,Callan:1994py,Kabat:1995eq,Frolov:1998vs}.

Several properties of black holes that are usually considered strange may be
reinterpreted in light of this interpretation. The area scaling of the entropy or of
other thermodynamic quantities is a property that is generic to spatially
entangled systems. In an entangled state, the entanglement entropy of the two
spatial regions must be equal, and this is suggestive that it be proportional to
the mutual surface area. A more thorough analysis shows that the area dependence
comes from the properties of short range correlations near the space boundary
\cite{Srednicki,Holzheyetal,Casini:2003ix,AreaScaling,Cramer:2005mx}.

If there is an interaction term in the Hamiltonian of the two wedges
(which may come from the kinetic terms for the fields when putting
spacetime on a lattice), then the entangled nature of the vacuum
will become physically relevant: when restricting an observer to one
wedge, information will seem to ``leak'' into the other wedge, and
so the evolution will be non-unitary. This argument may be made more
precise and will appear in \cite{tbp}.

\begin{acknowledgments}

The research of M.~B.~E. was supported in part by the National Science Foundation
under Grant No. PHY99-07949. A.Y. acknowledges the Kreitman foundation fellowship
and the Kavli Institute for Theoretical Physics where part of this research has
been carried out. We would like to thank O. Aharony, S. de Alwis, N. Itzhaki, M.
Kleban, J. Maldacena, D. Oaknin, E. Rabinovici, J. Simon and D. Terno for useful
discussions. We thank S. de Alwis, G. Lifschytz, and E. Rabinovici for comments on
the manuscript.
\end{acknowledgments}



\begin{thebibliography}{10}

\bibitem{Bekenstein}
J.~D. Bekenstein,
\newblock Phys. Rev. {\bf D7}, 2333 (1973).

\bibitem{Hawking}
S.~W. Hawking,
\newblock Commun. Math. Phys. {\bf 43}, 199 (1975).

\bibitem{BekEBounds}
J.~D. Bekenstein,
\newblock Phys. Rev. {\bf D23}, 287 (1981).

\bibitem{HoltHooft}
G.~'t~Hooft,
\newblock gr-qc/9310026.

\bibitem{HolSusskind}
L.~Susskind,
\newblock J. Math. Phys. {\bf 36}, 6377 (1995), [hep-th/9409089].

\bibitem{Bousso}
R.~Bousso,
\newblock Rev. Mod. Phys. {\bf 74}, 825 (2002), [hep-th/0203101].

\bibitem{AdSCFTreview}
    J.~M.~Maldacena,
    \newblock Adv. Theor. Math. Phys. {\bf 2}, 231 (1998),
    [hep-th/9711200].
For a review see
    O.~Aharony, S.~S. Gubser, J.~M. Maldacena,
    H.~Ooguri and Y.~Oz,
    \newblock Phys. Rept. {\bf 323}, 183 (2000), [hep-th/9905111].

\bibitem{StromVafa}
A.~Strominger and C.~Vafa,
\newblock Phys. Lett. {\bf B379}, 99 (1996), [hep-th/9601029].

\bibitem{MalStrWit}
J.~M. Maldacena, A.~Strominger and E.~Witten,
\newblock JHEP {\bf 12}, 002 (1997), [hep-th/9711053].

\bibitem{Bring}
M.~Cyrier, M.~Guica, D.~Mateos and A.~Strominger,
\newblock Phys. Rev. Lett. {\bf 94}, 191601 (2005), [hep-th/0411187].

\bibitem{Peetetal}
S.~S. Gubser, I.~R. Klebanov and A.~W. Peet,
\newblock Phys. Rev. {\bf D54}, 3915 (1996), [hep-th/9602135].

\bibitem{SenBHstring}
A.~Sen,
\newblock Mod. Phys. Lett. {\bf A10}, 2081 (1995), [hep-th/9504147].

\bibitem{Mathur}
O.~Lunin and S.~D. Mathur,
\newblock Phys. Rev. Lett. {\bf 88}, 211303 (2002), [hep-th/0202072].

\bibitem{Mathur:2004sv}
  S.~D.~Mathur,
  arXiv:hep-th/0401115.




\bibitem{KayWald}
B.~S. Kay and R.~M. Wald,
\newblock Phys. Rept. {\bf 207}, 49 (1991).

\bibitem{Maldacena:2001}
J.~M. Maldacena,
\newblock JHEP {\bf 04}, 021 (2003), [hep-th/0106112].

\bibitem{SusskindUglum}
  L.~Susskind and J.~Uglum,
  Phys.\ Rev.\ D {\bf 50}, 2700 (1994)
  [arXiv:hep-th/9401070].

\bibitem{Callan:1994py}
  C.~G.~Callan and F.~Wilczek,
  Phys.\ Lett.\ B {\bf 333}, 55 (1994)
  [arXiv:hep-th/9401072].

\bibitem{Kabat:1995eq}
  D.~Kabat,
  Nucl.\ Phys.\ B {\bf 453}, 281 (1995)
  [arXiv:hep-th/9503016].


\bibitem{Frolov:1998vs}
  V.~P.~Frolov and D.~V.~Fursaev,
  Class.\ Quant.\ Grav.\  {\bf 15}, 2041 (1998)
  [arXiv:hep-th/9802010].

\bibitem{Israel}
W.~Israel,
\newblock Phys. Lett. {\bf A57}, 107 (1976).

\bibitem{KabStr}
D.~Kabat and M.~J. Strassler,
\newblock Phys. Lett. {\bf B329}, 46 (1994), [hep-th/9401125].

\bibitem{Jacobson}
T.~Jacobson,
\newblock Phys. Rev. {\bf D50}, 6031 (1994), [gr-qc/9407022].


\bibitem{Thorne:1986iy}
  K.~S.~.~Thorne, R.~H.~.~Price and D.~A.~Macdonald,
  { \em Black Holes: The Membrane Paradigm},
  Yale University Press, 1986.


\bibitem{Schmidt}
E.~Schmidt,
\newblock Math. Ann. {\bf 63}, 433 (1907). For a more modern
discussion, see, e.g. M.~A. Nielsen and I.~L. Chuang, {\em Quantum
computation and quantum information}, Cambridge University Press,
UK, 2000, (Chapter 2.5).

\bibitem{Feynmanstatmech}
R.~P. Feynman,
\newblock {\em Statistical Mechanics: A Set of Lectures} (Perseues Publishing,
  1998).

\bibitem{TFD}
Y.~Takahashi and H.~Umezawa,
\newblock Collective phenomenon {\bf 2}, 55 (1975).



\bibitem{deAlwis1}
S.~P. de~Alwis and N.~Ohta,
\newblock hep-th/9412027.

\bibitem{deAlwis2}
S.~P. de~Alwis and N.~Ohta,
\newblock Phys. Rev. {\bf D52}, 3529 (1995), [hep-th/9504033].

\bibitem{'tHooft:1985}
G.~'t~Hooft,
\newblock Nucl. Phys. {\bf B256}, 727 (1985).

\bibitem{Jerusalemgroup}
A.~Giveon, A.~Konechny, E.~Rabinovici and A.~Sever,
\newblock JHEP {\bf 07}, 076 (2004), [hep-th/0406131].

\bibitem{BTZ}
M.~Banados, M.~Henneaux, C.~Teitelboim and J.~Zanelli,
\newblock Phys. Rev. {\bf D48}, 1506 (1993), [gr-qc/9302012].

\bibitem{Townsend}
P.~K. Townsend,
\newblock gr-qc/9707012.

\bibitem{Shortest}
R.~Brustein, D.~Eichler, S.~Foffa and D.~H. Oaknin,
\newblock Phys. Rev. {\bf D65}, 105013 (2002), [hep-th/0009063].

\bibitem{beb1}
J.~D. Bekenstein,
\newblock Int. J. Theor. Phys. {\bf 28}, 967 (1989).

\bibitem{gsl}
R.~Brustein,
\newblock Phys. Rev. Lett. {\bf 84}, 2072 (2000), [gr-qc/9904061].

\bibitem{gslstring}
R.~Brustein, S.~Foffa and R.~Sturani,
\newblock Phys. Lett. {\bf B471}, 352 (2000), [hep-th/9907032].

\bibitem{Brustein:2001di}
R.~Brustein, S.~Foffa and G.~Veneziano,
\newblock Phys. Lett. {\bf B507}, 270 (2001), [hep-th/0101083].

\bibitem{HawMalStr}
S.~Hawking, J.~M. Maldacena and A.~Strominger,
\newblock JHEP {\bf 05}, 001 (2001), [hep-th/0002145].

\bibitem{Branepuzzle}
 R.~Brustein, D.~Eichler and S.~Foffa,
  Phys.\ Rev.\ D {\bf 71}, 124015 (2005)
  [arXiv:hep-th/0404230].

\bibitem{BarbonRab}
J.~L.~F. Barbon and E.~Rabinovici,
\newblock JHEP {\bf 11}, 047 (2003), [hep-th/0308063].

\bibitem{PRK}
M.~Kleban, M.~Porrati and R.~Rabadan,
 M.~Kleban, M.~Porrati and R.~Rabadan,
  JHEP {\bf 0410}, 030 (2004)
  [arXiv:hep-th/0407192].


\bibitem{Witten9803}
E.~Witten,
\newblock Adv. Theor. Math. Phys. {\bf 2}, 505 (1998), [hep-th/9803131].






\bibitem{Srednicki}
M.~Srednicki,
\newblock Phys. Rev. Lett. {\bf 71}, 666 (1993), [hep-th/9303048].


\bibitem{Holzheyetal}
C.~Holzhey, F.~Larsen and F.~Wilczek,
\newblock Nucl. Phys. {\bf B424}, 443 (1994), [hep-th/9403108].

\bibitem{Casini:2003ix}
  H.~Casini,
  Class.\ Quant.\ Grav.\  {\bf 21}, 2351 (2004)
  [arXiv:hep-th/0312238].

\bibitem{AreaScaling}
  A.~Yarom and R.~Brustein,
  Nucl.\ Phys.\ B {\bf 709}, 391 (2005)
  [arXiv:hep-th/0401081].


\bibitem{Cramer:2005mx}
  M.~Cramer, J.~Eisert, M.~B.~Plenio and J.~Dreissig,
  arXiv:quant-ph/0505092.




\bibitem{tbp}
R.~Brustein, M.~Einhorn and A.~Yarom,
\newblock to appear.

\end{thebibliography}
\end{document}